# Assembly, Integration and Verification Activities for a 2U CubeSat, EIRSAT-1

Sarah Walsh, David Murphy, Maeve Doyle, Joseph Thompson, Rachel Dunwoody, Masoud Emam, Jessica Erkal, Joe Flanagan, Gianluca Fontanesi, Andrew Gloster, Joe Mangan, Conor O'Toole, Favour Okosun, Rakhi Rajagopalan Nair, Jack Reilly, Lána Salmon, Daire Sherwin, Paul Cahill, Daithí de Faoite, Umair Javaid, Lorraine Hanlon, David McKeown, William O'Connor, Kenneth Stanton, Alexei Ulyanov, Ronan Wall and Sheila McBreen
University College Dublin
Dublin, Ireland
sarah.walsh.2@ucdconnect.ie

*Abstract*—The Educational Irish Research Satellite, EIRSAT-1, is a project developed by students at University College Dublin that aims to design, build, and launch Ireland's first satellite. EIRSAT-1 is a 2U CubeSat incorporating three novel payloads; GMOD, a gamma-ray detector, EMOD, a thermal coating management experiment, and WBC, a novel attitude control algorithm. The EIRSAT-1 project is carried out with the support of the Education Office of the European Space Agency, under the educational Fly your Satellite! programme.

The Assembly, Integration and Verification (AIV) plan for EIRSAT-1 is central to the philosophy and the development of the spacecraft. The model philosophy employed for the project is known as the 'prototype' approach in which two models of the spacecraft are assembled; an Engineering Qualification Model (EQM) and a Flight Model (FM). The payloads, GMOD and EMOD, and the Antenna Deployment Module (ADM) platform element warrant a Development Model (DM) in addition to an EQM and a FM, as they have been designed and developed in-house. The engineering qualification model serves as both a FlatSat for electrical integration testing and as a representative model for testing of software code, patching and operational decisions during the active mission. The EQM is tested to qualification levels and durations during the environmental test campaign. The flight model contains the flight versions of the payloads, ADM platform element and the procured hardware elements. It undergoes acceptance level testing and it is the final spacecraft to be delivered to ESA for flight.

After successful completion of the Critical Design Review (CDR) and Ambient Test Readiness Review (ATRR) phases of the project, the EQM of EIRSAT-1 will be assembled and integrated in University College Dublin. After assembly and integration of the EQM, the project will begin the ambient test campaign, in which the EQM undergoes ambient functional and mission testing. This work details the preparation and execution of the assembly, integration, and verification activities of EIRSAT-1 EQM.

I. INTRODUCTION

The Educational Irish Research Satellite [1], EIRSAT-1, is a student led project which aims to design, build, launch and operate Ireland's first satellite. Selected in 2016 by ESA Education's Fly Your Satellite! (FYS) programme, EIRSAT-1 is a 2U CubeSat (Fig. 1) being developed at University College Dublin (UCD). The mission carries three payloads; a gamma-ray detector, the Gamma-Ray Module (GMOD), a thermal coating management experiment, the ENBIO Module (EMOD), and an attitude control algorithm, Wave Based Control (WBC) [2].

The Assembly, Integration and Verification (AIV) plan describes the overall Assembly, Integration and Testing (AIT) activities and the Verification Plan (VP) of the project. It is used to prepare the Assembly and Integration Procedure (AIP) and the Test Specification and Test Procedures (TSTP). The AIV plan of EIRSAT-1 describes the 'prototype' model philosophy that has been adapted for the project, the main activities in the assembly and integration process, and the tests carried out at subsystem and system level. The overall objective of verification is to demonstrate that the deliverable spacecraft meets the specified requirements. It demonstrates the qualification of the design and it confirms that the overall system is able to fulfil the mission requirements.

This paper details the AIV activities performed in preparation for the launch EIRSAT-1. Section II describes the model philosophy for the project with particular focus on the system level and on the in-house developed elements. Section III discusses the preparation for ambient testing. Section IV details the ambient test campaign for EIRSAT-1 while Section V details the environmental tests that will be performed.

II. MODEL PHILOSOPHY

The model philosophy defines the optimum number and the characteristics of the physical models produced to achieve confidence in the product, while weighing both the costs and risk involved. The model philosophy of EIRSAT-1 follows a 'prototype' approach. The prototype approach offers the project minimum risk, completion of qualification activities prior to acceptance, and the capability to use the Engineering Qualification Model as an integration spare at system level. For EIRSAT-1, the approach is covered in three main steps:

- Development Model (DM)/Element Testing.
- Engineering Qualification Model (EQM).
- Flight Model (FM).

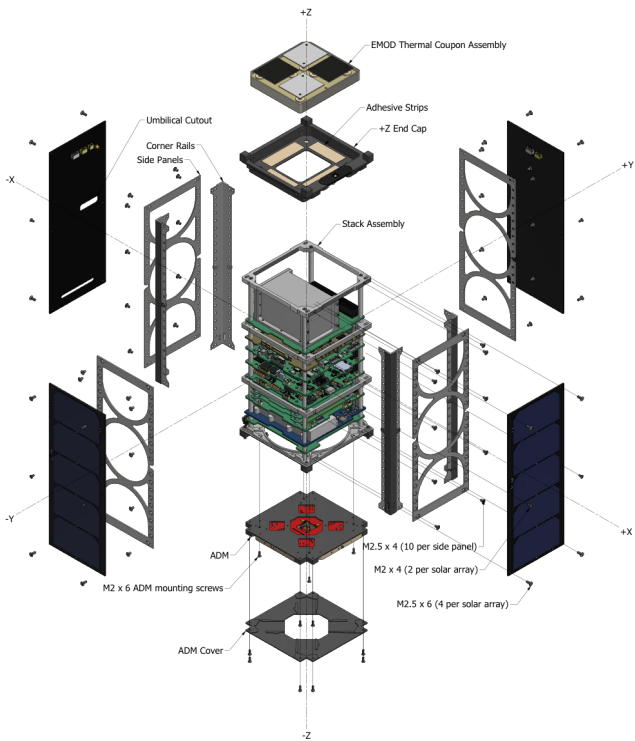

Fig. 1. Exploded view of EIRSAT-1.

Development models are produced for all elements of the spacecraft that are new developments and are not off the shelf. These elements are being developed in UCD and are most notably the payloads, GMOD, EMOD, and WBC, and the Antenna Deployment Module (ADM). These models allow confirmation of the design feasibility of the element. They are designed in an iterative process whereby each element DM is designed, assembled and tested at unit level until all functional, electrical, and software requirements are in agreement with verification objectives. For elements that are commercial off the shelf (COTS), two one-off amounts are procured to have one set each for the spacecraft EQM and FM.

*A. Spacecraft*

Two models exist at the spacecraft level – the Engineering Qualification Model and the Flight Model. The EQM is made up of the EQMs of the payloads, ADM platform element and the procured subsystems. The model serves both as a FlatSat for electrical integration testing and also as a representative model for testing of software code and for operational decisions during the active mission. The model provides the functional qualification of the design and the interfaces between subsystems. It undergoes testing to qualification levels and durations. Any feedback from qualification testing of the EQM is taken into account when manufacturing the FM. The FM consists of the flight COTS subsystems, FM payloads and FM ADM. It is the end product that is intended for flight and is subject to functional and environmental acceptance level testing before delivery to ESA.

The design of both the EQM and FM is identical, within the project constraints. Producing both an EQM and FM at system level has significant costs for the project but hugely reduces the risk involved. To save on cost, the EQM does not include solar cells. However, suitable custom boards have been produced to mimic the magnetorquers and Coarse Sun Sensors (CSS) of the ADCS subsystem to allow for testing on the EQM.

*B. GMOD Payload*

GMOD is an experiment designed to detect cosmic gamma-ray phenomena such as Gamma-Ray Bursts (GRBs). The payload can be considered in two parts: the detector assembly and the motherboard. The detector assembly contains a $CeBr_3$ scintillator, a 4 x 4 Silicon Photomultiplier (SiPM) array and an Application Specific Integrated Circuit (ASIC). The motherboard has been designed in-house and contains the supporting electronics of the detector assembly.

The GMOD detector assembly will have an EQM and a FM, which will be identical. Therefore, the EQM can serve as a flight-spare or on-ground debugging tool. The mature design of the detector assembly does not warrant a DM. The EQM will undergo environmental testing at unit-level before being integrated with the full spacecraft EQM. An additional model will be produced to undergo shock testing, verifying the functionality of the payload after encountering the loads of the launcher. While the detector assembly does not require a DM, the low cost associated with the motherboard components meant that producing development models significantly reduced the risk associated with the GMOD payload. The GMOD motherboard underwent a design process in which multiple development models were manufactured until the final design of the board was accepted. The final DM of the motherboard (DM3) is the design on which the EQM and FM are based.

*C. EMOD Payload*

EMOD is an experimental payload designed to demonstrate and test the performance of two thermal spacecraft coatings developed by ENBIO Ltd. The experiment consists of the Thermal Coupon Assembly (TCA), which is affixed to the +Z face of the spacecraft, and the motherboard. The TCA accommodates four 'thermal coupons' which have been treated with SolarBlack or SolarWhite thermal control coating. The temperature of the coupons will be monitored throughout the lifetime of the EIRSAT-1 mission.

The EMOD TCA has a DM, an EQM, and a FM. The DM was developed to conduct adhesion and outgassing tests of the ENBIO adhesive primer [3]. The EQM and the FM shall be identical and so the EQM can act as a flight-spare or on-ground debugging tool. The TCA EQM will undergo environmental testing at unit level to verify the design and workmanship of the assembly. Like GMOD, the design of the EMOD motherboard was finalised after an iterative process. The final DM (DM4) of the motherboard is the design for the EQM and FM.

*D. Antenna Deployment Module*

The ADM [4] holds two dipole antennae and is attached to the –Z end cap of EIRSAT-1. The deployment mechanism uses pairs of melt-lines that hold spring loaded doors in a closed position. The lines are burnt during deployment to allow the antennae to uncoil into their straight deployed configuration.

The ADM model philosophy follows that of GMOD and EMOD consisting of a DM, an EQM, and a FM. The ADM has undergone extensive design and testing due to its significance in performing the mission. A mechanical prototype (DM) of the element was manufactured and was tested under both ambient and environmental conditions to verify the mechanical functionality of the design. In parallel with the development of the mechanical design, the electrical design and the efficiency of the antenna elements in terms of the communications subsystem have been configured and tested. At DM level, two parallel designs were implemented on Printed Circuit Boards (PCBs) and the radiation pattern of both were tested. A PCB design was chosen after testing as the final design of the ADM PCB, which will implemented for the EQM and the FM.

### III. AMBIENT TEST READINESS REVIEW

The Ambient Test Readiness Review (ATRR) for EIRSAT-1 began in June 2018. The review process involved the preparation of the Assembly and Integration Plan and of the Test Specification and Test Plan (TSTP) for all tests to be carried out during the Ambient Test Campaign (ATC). These procedures were subject to a review process from ESA and review item discrepancies (RIDs) were issued to the team. After the closure of the RIDs, the EQM of EIRSAT-1 was assembled.

*A. Assembly and Integration Plan*

The assembly and integration plan of EIRSAT-1 defines the step-by-step procedure that is followed during assembly of the EQM and the FM. The requirements and steps outlined in the document must be followed in detail to ensure that the spacecraft is assembled correctly. The main steps taken to assemble and integrate both the EQM and FM of EIRSAT-1 are illustrated in Figure 2. The assembly procedure is undertaken by two assigned operators while a product assurance (PA) team member is present to record the process. Operators must ensure that all ground support equipment, tools and consumables listed in the AIP that are required during the procedure are present in the cleanroom. All components and parts of the spacecraft used for assembly must also be present and inspected for any defects.

Before the assembly and integration of the main spacecraft begins, the deployment switches are installed to the corner rails while thermocouples are installed to the subsystem PCBs. Thirteen thermocouples are used during the environmental test campaign to monitor the internal temperatures of certain subsystems. These thermocouples are installed to the spacecraft using a combination of kapton and aluminium tape for the EQM while a non-conducting epoxy glue will be used for the FM. The assembly and integration process begins from the bottom of the stack assembly with the –Z end cap. The four support rods are attached to the end cap and an orthogonality check of the rods is performed using a support interface rib. Next, the PCBs are placed onto the stack in the order of the Z-axis magnetorquer (MTQ), Communications (TT&C), Battery, Electrical Power System (EPS), On Board Computer (OBC), Attitude Determination and Control System (ADCS), EMOD motherboard, and finally, the GMOD motherboard and detector assembly. Two support brackets are placed in the stack, one between the EPS and the OBC while the other is located between the EMOD and GMOD motherboards. All PCBs and support brackets are supported by 4 spacers, one on each of the four support rods, and the PCBs are electrically connected by the 104-pin header at the +Y face of the stack. As each PCB is placed on the stack assembly, any harnesses for that PCB are connected and are routed through the stack. Finally, the upper support bracket is secured to the +Z ends of the support rods to complete the PCB stack assembly. The next step of the assembly process is to attach the side panels. The M2.5 screws that hold the side panels in place also fix the corner panels to the structure of the spacecraft, and so the corner rails must be temporarily attached to the PCB stack. The rails are attached by nylon screws which are inserted into the corner rail clearance holes at the –X and +X faces of the spacecraft. This allows the side panels to be mounted onto the –Y and +Y faces using the desired M2.5 Al-6082-T6 screws. Next, the EMOD payload is integrated to the spacecraft and the +Z end cap is secured to the corner rails. The temporary nylon screws are removed from the rails so that the +X and –X side panels can be mounted to the spacecraft. Next, the ADM is attached by rotating the CubeSat by 90° in the integration stand so that the –Z end cap is now accessible. The outer cover of the ADM is removed and the ADM is secured to the –Z end cap by 6 M2 mounting screws. The outer cover is then replaced back onto the ADM. Finally, the solar panels are mounted onto the spacecraft. The order of the solar panel mounting is –Y, +X, +Y, and –X. The ordering of the solar panel attachment is to facilitate the routing of thermocouple cables that are attached to the back of the solar panel PCBs. All thermocouples which are placed throughout the spacecraft for use during environmental testing exit the spacecraft through a 50mm x 5mm cut out in the –X solar panel.

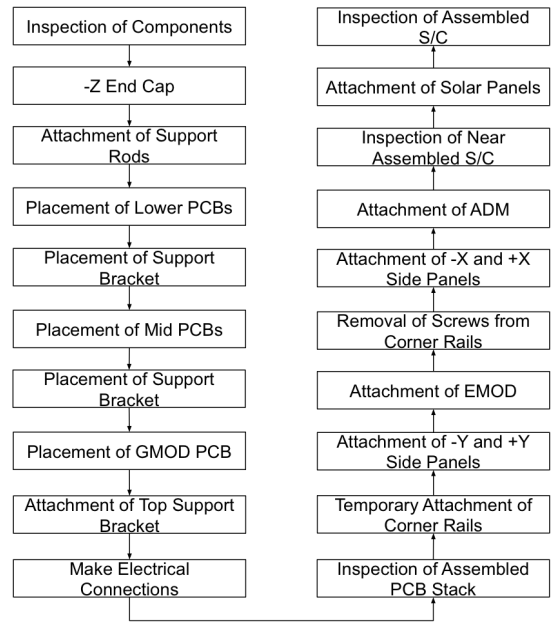

Fig. 2. Flow chart of the assembly and integration activites for EIRSAT-1.

Various inspection points are performed during the assembly process. These inspections occur once a central part of the spacecraft has been assembled, such as after the PCB stack assembly, after the side panels have been attached and finally, once the spacecraft has been fully assembled. These inspection points allow the operators to note any defects or anomalies during the assembly process to reduce the risk of failures during the test campaigns that follow.

*B. Mass and Dimension Verification*

After the assembly and integration of EIRSAT-1, the mass and the dimensions are measured to verify that the CubeSat is within the limits of mass and dimensions as required by the Fly Your Satellite Design Specification (FDS). When fully integrated, the EIRSAT-1 spacecraft is expected to be 2066g, compliant with the requirement of 2660g for a 2U CubeSat.

*C. Phase D Workshop*

In April 2019, a Phase D Workshop was held by ESA Education and FYS. The EIRSAT-1 team presented a representative model of the EQM which incorporated the EQMs of the Z-axis MTQ, TT&C, OBC, Battery, EPS, and structure subsystems; the DMs of the ADM, GMOD, and EMOD; and a dummy PCB for the ADCS subsystem. During the workshop, the mass and dimension verification of the model was performed. The mass was measured as 1681.7g, less than the total expected mass of EIRSAT-1 due to the scintillator crystal not being present in the GMOD detector assembly and a dummy PCB being present for the ADCS subsystem.

The dimensions verification was carried out for the CubeSat's width and height. The width of the model was measured at the top, middle and bottom of the main structure and at the +Z and –Z end cap rail standoffs for each of the –X, +X, –Y and +Y sides of the unit. All measured values were compliant with the requirement of $100.0 \pm 0.1$mm. The height of the spacecraft was measured for each of the four corner rails from the –Z to the +Z end cap rail standoffs. All measurements were compliant with the requirement of $227.0 \pm 0.2$ mm for a 2U CubeSat. A further verification of the dimensions of the CubeSat were carried out by insertion into a CubeSat deployer. An inspection verification of the model was performed during which the exterior of the CubeSat was inspected for any visible damage. During the inspection, any scratches to the exterior were noted.

After the workshop, the model of EIRSAT-1 was disassembled, following the disassembly procedure. This facilitated further testing of the mission software on the subsystem EQMs in a FlatSat configuration while the payloads and ADM platform element EQMs were qualified at subsystem level. EMOD and the ADM underwent vibration and thermal vacuum testing, while vibration and shock tests were carried out on GMOD. This allowed for further confidence to be obtained in the design and workmanship of these in-house developed subsystems before integration of the EQM.

IV. AMBIENT TEST CAMPAIGN

The ambient test campaign for EIRSAT-1 is planned to begin in October 2019. It is expected to last 3 months after which test reports (TRTPs) for all tests carried out during the campaign will be submitted to FYS! for review. All test activities to be carried out during the ambient test campaign have a separate Test Specification and Test Procedure. The Test Specification describes the test requirements applicable to a related test activity while the Test Procedure gives the directions for conducting the test. The documents are written with such level of detail that different test operators can obtain the same results when executing the procedure. The step-by-step procedure is used to conduct the test and is filled in throughout. It becomes the core section of the test report. The two main tests to be carried out during the ambient test campaign for EIRSAT-1 are the functional test and the mission test.

*A. Functional Test*

Functional testing verifies the complete function of the spacecraft, under the specified operating and environmental conditions and in all operational modes. Functional tests are carried out on both the EQM and the FM models of EIRSAT-1 and occur at the beginning and the end of various test activities such as thermal vacuum and vibration testing. The results of the functional tests carried out on the spacecraft during various stages of the test campaigns should be identical, within the test tolerances. The test includes testing mechanical functions, electrical functions and operational functions of the spacecraft. The functional test planned for EIRSAT-1 consists of the following:

- Visual inspection.
- Antenna deployment of all elements testing the primary and secondary melt-line deployment mechanisms.
- Testing of all operational modes and the CubeSat separation sequence including removal of Remove Before Flight (RBF) pin, deployment of activation switches, antenna deployment.
- Battery and EPS testing including battery charging via solar cells and charging tether cable, EPS protection functions, battery discharging, EPS trip recording, low battery boot up, and RBF pin insertion.
- Flashing of GMOD and EMOD firmware.
- GMOD functionality including burst detection, sensitivity to laboratory source, and processing of light curves and spectra.
- EMOD functionality including polling temperature data and varying the data acquisition rate.
- WBC functionality including testing of ADCS modes.
- Software reprogramming from the ground for any updates or bug fixes made to software.

*B. Mission Test*

The mission test ensures that the spacecraft is capable of performing the required mission when in-orbit, by simulating in-flight operations, sequences and hardware/software interfaces that will (or may) occur over the lifetime of the mission. This test includes the main, critical, and contingency operations that are foreseen for the mission. To perform this test, the entire mission profile is simulated as if in-flight, within the constraints of a ground-based test facility. Within the mission test, end-to-end testing will also ensure that the ground segment, as well as the space segment, is capable of operating as required for the success of the mission. EIRSAT-1's mission test will test the following mission scenarios:

- Deployment and the separation sequence including removal of RBF pin, deployment of activation switches, antenna deployment.
- Initial Acquisition of Signal (AOS) pass.
- Nominal pass, including a prolonged nominal pass.
- WBC mode entry for both a scheduled and requested telecommand.
- Automatic transition from WBC mode to safe mode in case of attitude parameters exceeding the predefined limits.
- Safe mode entry and exit.
- Hard reset of the spacecraft.
- Watchdog timer reboot after a prolonged period of silence from the ground station.
- Transmission inhibit in the event of a request to cease transmissions of the spacecraft for a period of time.
- Force mode change from the current operational mode to another.
- Power cycling of the payloads.
- Operating on a low battery level.
- Uploading and booting to a new software image.

## V. Environmental Test Campaign

The environmental test campaign for EIRSAT-1 will begin after successful review of all ambient test reports, which is planned for early 2020. During the environmental test campaign, the EQM is tested to qualification levels to verify that the design and manufacturing technique fulfill specification requirements. It accounts for severe hardware characteristics which can be present in a flight unit. The FM will be tested to acceptance levels to check the workmanship of the flight hardware and to avoid inducing additional stress on the unit that may lead to a failure during the mission.

### A. Vibration Test

The vibration test will be the first environmental test to be carried out at system level as it will be the first environment encountered during the mission. The vibration test aims to replicate the launch mechanical environment that EIRSAT-1 will experience. To give confidence in the design of the spacecraft before testing, modelling and analyses were carried out to assess the stress and displacement patterns of EIRSAT-1. The analysis provides the locations of the spacecraft and the vibration frequencies that should be monitored closely throughout the vibration test. The test will consist of a modal survey, random vibrations, and sinusoidal vibrations of all 3-axes of the CubeSat. EIRSAT-1 will be placed within a CubeSat deployer for the duration of the test as this will be its configuration for launch. Dummy masses will also be inside the deployer to replicate nearby CubeSats. Sensors will be placed on EIRSAT-1 during the test to measure the displacement and acceleration of the CubeSat.

Vibration testing will take place at the CubeSat Support Facility of the ESA Education Training Centre. The facility is equipped with a 20kN electro-dynamic shaker.

### B. Thermal Vacuum Test

In line with the launch sequence, the thermal vacuum test will occur after the vibration test. The test will ensure that the CubeSat is capable of fulfilling all functional requirements under various temperature conditions that will be encountered during flight. It also verifies the workmanship and thermal design of the system. The thermal vacuum test will consist of 1 cycle within the non-operational temperature range and 3 cycles within the operational temperature range of the CubeSat. It will also consist of dwell phases during which the temperature will be maintained at a constant value for approximately two hours. The temperature ranges for the test is defined by the thermal analysis.

Thermal vacuum testing of EIRSAT-1 will take place at the CubeSat Support Facility of the ESA Education Training Centre. The facility is equipped with a thermal vacuum chamber with a temperature range of -60°C – +100°C and a vacuum limit of $10^{-6}$ mbar.

## Acknowledgments

We acknowledge all students who have contributed to EIRSAT-1. The EIRSAT-1 project is carried out with the support of the Education Office of the European Space Agency under the educational Fly Your Satellite! programme.

SW acknowledges support from the European Space Agency under PRODEX contract number 4000120713. DM acknowledges support from the Irish Research Council (IRC) under grant GOIPG/2014/453. JT acknowledges support from the IRC under grant GOIPG/2014/684. MD acknowledges support from the IRC under grant GOIP/2018/2564. JE, JR and RD acknowledges scholarships from the UCD School of Physics. CO'T acknowledges support from the IRC under grant GOIPG/2017/1031. LS acknowledges support from the IRC under grant GOIPG/2017/1525. JM and AU acknowledge support from Science Foundation Ireland under grant number 17/CDA/4723.